\documentclass[amsmath,amssymb,aps,prd,reprint,twocolumn,showkeys,showpacs]{revtex4-2}

\usepackage{xcolor}
\usepackage{amsmath}
\usepackage{graphicx}
\usepackage{dcolumn}
\usepackage{bm}
\usepackage{epstopdf}
\printfigures
\usepackage{epsfig}
\usepackage[]{quoting}
\usepackage{subfigure}
\usepackage{amssymb }
\usepackage{booktabs}

\begin{document}

\title{Physical aspects of ${f}(R,G_{\mu \nu}T^{\mu \nu})$ modified gravity theories}

\author{Mihai Marciu}
\email{mihai.marciu@drd.unibuc.ro}
\affiliation{Faculty of Physics, University of Bucharest, 405 Atomi\c{s}tilor, POB MG-11, RO-077125, Bucharest-M\u{a}gurele, Romania}

\author{Dana Maria Ioan}
\email{idana91@yahoo.com}
\affiliation{Faculty of Physics, University of Bucharest, 405 Atomi\c{s}tilor, POB MG-11, RO-077125, Bucharest-M\u{a}gurele, Romania}

\begin{abstract}
The paper extends basic Einstein--Hilbert action by adding a newly proposed invariant constructed from a specific contraction between the Einstein tensor and the energy momentum tensor, encoding a non--minimal coupling between the space--time geometry and the matter fields. The fundamental Einstein--Hilbert action is extended by considering a generic function  ${f}(R,G_{\mu \nu}T^{\mu \nu})$ which is further decomposed into its main constituents, a geometric component which depends on the scalar curvature, and a second element embedding the interplay between geometry and matter fields. Specific cosmological models are established at the level of background dynamics, based on particular couplings between the matter energy--momentum tensor and the Einstein tensor. After deducing the resulting field equations, the physical aspects for the cosmological model are investigated by employing a dynamical system analysis for various coupling functions. The investigation showed that the present model is compatible with different epochs in the evolution of our Universe, possible explaining various late time historical stages.  
\end{abstract}

\maketitle

\newpage 

\section{Introduction}
\par
In the recent years the cosmological theories have developed intensively due to various astrophysical probes \cite{aghanim2020planck, akrami2020planck}, adding new intriguing questions to modern physics. In the cosmological theories, one of the greatest challenges is related to the accelerated expansion of the known Universe \cite{frieman2008dark, yoo2012theoretical}, an interesting phenomenon which is associated to the evolution of the Universe at the large scale structure \cite{Li:2012dt, Copeland:2006wr, arun2017dark, brax2017makes, peebles2003cosmological}. Another key problem in cosmology is related to the dark matter phenomenon \cite{bertone2018history, garrett2011dark, freese2017status}, having various ramifications at the level of galactic dynamics, being more local. All of these aspects suggest that we are far from understanding the nature and evolution of the Universe as a whole, leaving the door open for new theoretical directions and cosmological models \cite{cervantes2011cosmology}. From an astrophysical point of view, various observational studies can probe the nature and properties of the accelerated expansion \cite{huterer2017dark, frieman2008dark, albrecht2007evaluating, huterer2001probing, abbott2019cosmological, wolf2020standard, elvin2018dark}. This phenomenon represents a curious aspect related to the fundamental ingredients of the Universe \cite{dark2016dark}, having various ramifications in science and technology. 
\par 
The simplest possible theoretical model related to the origin and characteristics of the accelerated expansion is the $\Lambda$CDM scenario \cite{Copeland:2006wr}, a proposal which cannot explain various fundamental aspects of this phenomenon \cite{Bahamonde:2017ize}. In order to solve the dark energy problem, various theoretical ramifications have emerged \cite{nojiri2007introduction}. In the modified gravity theories, the fundamental Einstein--Hilbert action is extended, by adding specific invariants which can be associated to the origin of the accelerated expansion \cite{bamba2012dark, Marciu:2021rdl, Marciu:2020ysf, Bahamonde:2019urw}. One of the first model is the $f(R)$ scenario \cite{nojiri2006modified}, a theoretical model which is based on the scalar curvature, extending the Einstein--Hilbert action in a more natural approach \cite{nojiri2011unified}. The modified gravity theories based on the $f(R)$ invariant have been studied intensively in the past decades \cite{nojiri2011unified, nojiri2008modified, cognola2005one, de2010f, sotiriou2010f, sebastiani2015f}. In different approaches, various alternative theories have been proposed, based on specific invariants in the corresponding action. For example, another origin of the dark energy phenomenon can be related to the $f(G)$ models \cite{nojiri2005modified, de2009construction, nojiri2017modified}, based on the Gauss--Bonnet invariant. In these theories, one might consider the extension towards the interplay between the matter lagrangian and the scalar curvature, applying a specific action $f(R,L_m)$ \cite{wang2012energy, harko2010f}. This particular model can be further generalized, by considering a scenario based on $f(R,T)$ \cite{harko2011f}, with $T$ the trace of the matter energy--momentum tensor \cite{alvarenga2013dynamics}. In the latter theory, we can consider matter and geometry on equal footing, a specific interplay which can trigger the accelerated expansion \cite{harko2011f}. All of these theories have been studied exhaustively in the last years \cite{moraes2017modeling, sharif2012thermodynamics, odintsov2013f, moraes2017simplest, shabani2014cosmological, zubair2016static, shabani2013f, rosa2021junction, zaregonbadi2016dark}, showing specific advantages over the $\Lambda$CDM scenario \cite{houndjo2012reconstruction}. The dark energy phenomenon can be also triggered by a scalar field embedded into action \cite{sola2017dynamical, Copeland:2006wr}, which can be minimally or non--minimally coupled \cite{Bahamonde:2017ize} in the form of quintessence \cite{PhysRevD.79.103005}, phantom \cite{PhysRevLett.91.071301}, or quintom scenarios \cite{CAI20101, Marciu:2019cpb, Marciu:2018oks, Marciu:2016aqq, Marciu:2020hpk,marciu2020dynamical,marciu2021dynamics}. In the modified gravity theories, a particular invariant based on the contraction between the matter energy--momentum tensor and the Ricci tensor $R_{\mu\nu}T^{\mu\nu}$ has been considered \cite{odintsov2013f, haghani2013further}, taking into account a possible interplay between the space--time geometry and the fundamental characteristics of the matter component, embedded into the specific form of the energy--momentum tensor. From a theoretical perspective, such an approach is viable and can lead to different effects \cite{universe8020062, YOUSAF2020100527, PhysRevD.99.124027, Sharif:2018sie, Abchouyeh:2020vfh, Sharif:2018tyi, SHARIF2021179, YOUSAF2020100581, PhysRevD.91.104003, Kaczmarek2021}. From this, the extension towards an invariant based on the interplay between the Einstein tensor and the energy--momentum tensor occurs naturally  \cite{https://doi.org/10.48550/arxiv.2212.03821} and can lead to a viable modified gravity theory. 
\par 
In a recent study \cite{https://doi.org/10.48550/arxiv.2212.03821}, the authors have investigated the possible interplay between the Einstein tensor and the matter energy--momentum tensor, considering also various specific contractions of the derivatives. The study showed the viability of such a model, taking into account a perturbation analysis by considering a linear model in the proposed action. Taking into account the possible coupling between the matter energy--momentum tensor and the Einstein tensor, we have further extended the corresponding model, by adding a general function which depends on the interplay between the matter and geometry, embedded into the Lagrangian. In order to make the model more general, we have also generalized the fundamental geometric part, embedding a specific component which further depends on the scalar curvature. Hence, the present model is further decomposed into a pure geometric part which depends on the scalar curvature, and a second component which depends on the new invariant, based on the interplay between matter and geometry, embedding the contraction between the energy--momentum tensor and the Einstein tensor. 
\par 
The paper is organized as follows: in Sec.~\ref{sec:adoua} we briefly describe the cosmological model and the corresponding field equations. Then, in Sec.~\ref{sec:atreia}, by employing the dynamical system analysis we investigate the generic model which takes into account the curvature and the matter couplings with the Einstein tensor. Furthermore, in Sec.~\ref{sec:apatra} a specific subclass is investigated, where the $f(R)$ part corresponds to the basic Einstein--Hilbert term. In Sec.~\ref{sec:acincea} the last proposed model is investigated, where the matter--geometry function is represented by an exponential model. Lastly, in Sec.~\ref{sec:concluzii} we summarize the most important results and briefly present the concluding remarks.

\section{The action and the corresponding field equations}
\label{sec:adoua} 
In this section we shall briefly discuss the main elements corresponding to the present model, deducing the gravitational field equations in a cosmological context. In what follows we shall propose a model described by the following action:
\begin{equation}
	S=S_m+\int d^4x \sqrt{-\tilde{g}} \big[f(R)+g(\phi) \big],
\end{equation}
where we have added to the $f(R)$ part \cite{RevModPhys.82.451} a new generic function which is further based on specific contractions between the Einstein tensor and the energy--momentum tensor \cite{https://doi.org/10.48550/arxiv.2212.03821}, 
\begin{equation}
	\phi=G_{\mu \nu} T^{\mu \nu}.
\end{equation}
Here the energy--momentum tensor is defined in the usual manner,
\begin{equation}
	T_{\mu \nu}=-\frac{2}{\sqrt{-\tilde{g}}}\frac{\delta(\sqrt{-\tilde{g}}L_m)}{\delta \tilde{g}^{\mu \nu}},
\end{equation}
with $L_m$ the Lagrangian for the matter sector. Before proceeding to the derivation of the Einstein field equations, we introduce new elements which shall be considered. The trace of the energy--momentum tensor is defined as: $T=T^{\mu \nu}\tilde{g}_{\mu \nu}$. The variation of the energy--momentum tensor with respect to the inverse metric is equal to the following relation \cite{https://doi.org/10.48550/arxiv.2212.03821}:
\begin{equation}
	\frac{\delta T_{\alpha \beta}}{\delta \tilde{g}^{\mu \nu}}=\frac{\delta \tilde{g}_{\alpha \beta}}{\delta \tilde{g}^{\mu \nu}}L_m+\frac{1}{2}\tilde{g}_{\alpha \beta}L_m \tilde{g}_{\mu \nu}-\frac{1}{2}\tilde{g}_{\alpha\beta}T_{\mu\nu}-2\frac{\partial^2 L_m}{\partial \tilde{g}^{\mu\nu} \partial \tilde{g}^{\alpha\beta}}.
\end{equation}
Next, in our computations we shall consider the following contraction \cite{https://doi.org/10.48550/arxiv.2212.03821}:
\begin{multline}
	\Sigma_{\mu\nu}=G^{\alpha\beta}\frac{\delta T_{\alpha\beta}}{\delta \tilde{g}^{\mu\nu}}=-G_{\mu\nu}L_m+\frac{1}{2}G^{\alpha\beta}\tilde{g}_{\alpha\beta}(\tilde{g}_{\mu\nu}L_m-T_{\mu\nu})
	\\-2 G^{\alpha\beta}\frac{\delta^2 L_m}{\delta \tilde{g}^{\mu\nu} \delta \tilde{g}^{\alpha\beta}}.
\end{multline}
For the $f(R)$ part, the variation of the corresponding action with respect to the inverse metric leads to the associated energy momentum tensor \cite{RevModPhys.82.451},
\begin{equation}
	T_{\mu\nu}^{f(R)}=\tilde{g}_{\mu\nu}f(R)-2 R_{\mu\nu}f_R+2\nabla_{\mu}\nabla_{\nu}f_R-2 \tilde{g}_{\mu\nu} \Box f_R,
\end{equation} 
where we have introduced the derivative with respect to the scalar curvature, $f_R=\frac{\partial f(R)}{\partial R}$.
In the case of $g(\phi)$ component, we have obtained the following energy--momentum tensor \cite{https://doi.org/10.48550/arxiv.2212.03821},
\begin{multline}
	T_{\mu\nu}^{g(\phi)}=\tilde{g}_{\mu\nu}g(\phi)+g_{,\phi}T R_{\mu\nu}-2 g_{,\phi} G_{\nu}^{\beta}T_{\mu\beta}-2 g_{,\phi} G_{\mu}^{\alpha}T_{\nu\alpha}
	\\-g_{,\phi}R T_{\mu\nu}-\Box (g_{,\phi} T_{\mu\nu})+\nabla_{\alpha}\nabla_{\mu}(g_{,\phi}T_{\nu}^{\alpha})+\nabla_{\alpha}\nabla_{\nu}(g_{,\phi}T_{\mu}^{\alpha})
	\\-\tilde{g_{\mu\nu}}\nabla_{\alpha}\nabla_{\beta}(g_{,\phi} T^{\alpha\beta})+\tilde{g_{\mu\nu}}\Box(g_{,\phi}T)-\nabla_{\mu}\nabla_{\nu}(g_{,\phi}T)-2g_{,\phi}\Sigma_{\mu\nu},
\end{multline}
with $g_{,\phi}$ the derivative with respect to the $\phi$ invariant, i.e.
\begin{equation}
	g_{,\phi}=\frac{d g(\phi)}{d \phi}.
\end{equation}
If we further apply the principle of least action we obtain the final Einstein--like equation \cite{RevModPhys.82.451},
\begin{equation}
    T_{\mu\nu}^{f(R)}+T_{\mu\nu}^{g(\phi)}+T_{\mu\nu}^{m}=0,
\end{equation}
which leads to the conservation relation,
\begin{equation}
	\nabla^{\mu}\big[T_{\mu\nu}^{f(R)}+T_{\mu\nu}^{g(\phi)}+T_{\mu\nu}^{m}\big]=0,
\end{equation}
also known as the continuity equation.
\par 
Next, we shall consider the following cosmological context associated to the FLRW model described by the metric:
\begin{equation}
	ds^2=-dt^2+a(t)^2\delta_{ij}dx^idx^j, i,j=1,2,3.
\end{equation}
In this case we take into account a universal scale factor $a$ which depends on cosmic time. Then, we define the Hubble parameter in the usual manner, $H=\frac{\dot{a}}{a}$, where the dot denotes differentiation with respect to the cosmic time. The energy momentum tensor for the barotropic matter fluid is the following:
\begin{equation}
	T_{\mu\nu}=(\rho_m+p_m)u_{\mu}u_{\nu}+p_m g_{\mu\nu},
\end{equation} 
with $\rho_m$ the density, $p_m$ the pressure, connected through a barotropic equation of state of the form: $p_m=w_m\rho_m$. In the expression of the energy--momentum tensor we have embedded the 4-velocity $u_{\mu}=\delta_\mu^0$. Furthermore, for the computations we shall consider $L_m=p_m$. In what follows we shall neglect the pressure for the matter sector, taking into account a theoretical scenario corresponding to a non--relativistic fluid without pressure. In this cosmological context we arrive at the following modified Friedmann relations \cite{https://doi.org/10.48550/arxiv.2212.03821}:
\begin{equation}
	\label{frconstraint}
	f(R)-6 f_R(\dot{H}+H^2)+6 H \dot{f_R}=\rho_m-g(\phi)-6 \rho_m g_{,\phi} \dot{H},
\end{equation}

\begin{equation}
	\label{acceleration}
	f(R)-2 f_R(\dot{H}+3 H^2)+2 \ddot{f_R}+4 H \dot{f_R}=-p_{\phi},
\end{equation}

\begin{multline}
		p_{\phi}=g(\phi)-2 g_{,\phi}(\rho_m(3 H^2+\dot{H})+H \dot{\rho_m})
		\\ -6 H^2 \rho_m \big[2 \rho_m \dot{H}+H \dot{\rho_m} \big] g_{,\phi\phi}.
\end{multline}

Then, we can define the effective(total) equation of state as:
\begin{equation}
	w_{tot}=-1-\frac{2}{3}\frac{\dot{H}}{H^2}.
\end{equation}
For the FLRW model without pressure the matter--geometry invariant acquires the following value, $\phi=3 H^2 \rho_m$, while the scalar curvature is equal to $R=6(\dot{H}+2 H^2)$. As expected, the resulting field equations reduces to the fundamental Einstein equations if $g(\phi)=0$ and $f(R)=\frac{R}{2}$. In the same manner, the model describes the $f(R)$ theories of gravitation \cite{RevModPhys.82.451} in the absence of the interplay between the matter energy--momentum tensor and the Einstein tensor. Finally, we note that at the linear level $(g(\phi)=g_0\phi)$ the equations reduces to the relations presented in Ref.~\cite{https://doi.org/10.48550/arxiv.2212.03821}. As previously stated, the action presented in the present paper offers a generalization for the analysis presented in Ref.~\cite{https://doi.org/10.48550/arxiv.2212.03821}, extending the field equations in a generic manner.

\section{The phase space analysis for the general coupling function}
\label{sec:atreia} 
\par 
In this section we shall discuss the physical features in the case where the coupling functions are the following:
\begin{equation}
	f(R)=f_0 R^n,
\end{equation}
and 
\begin{equation}
	g(\phi)=g_0 \phi^m,
\end{equation}
with $f_0, g_0, n, m$ constant parameters. In order to study the cosmological model we have to introduce the following dimension--less variables:
\begin{equation}
	x=\frac{\dot{f_R}}{f_R H},
\end{equation}

\begin{equation}
	z=\frac{R}{6 H^2},
\end{equation}

\begin{equation}
	s=\frac{\rho_m}{6 f_R H^2},
\end{equation}

\begin{equation}
	u=\frac{g(\phi)}{6 f_R H^2}.
\end{equation}

Furthermore, we shall use the next non--independent variables:
\begin{equation}
	\varsigma=\frac{\ddot{R}}{H^4},
\end{equation}

\begin{equation}
	\Delta=\frac{\dot{\rho_m}}{f(R) H}.
\end{equation}

Considering the above definitions, we can rewrite the Friedmann constraint equation \eqref{frconstraint} in the following way:
\begin{equation}
	z(\frac{1}{n}-1)+1+x=s-u-2 u m z +4 u m,
\end{equation}
reducing the dimension of the autonomous system with one unit. Hence, we remain with the following autonomous system $\{z,s,u\}$. Considering the transformation to the e-folding number $N=log(a)$, we can approximate the dynamics of the cosmological system at the linear level, 
\begin{equation}
	\label{unu}
	\frac{dz}{dN}=z \left(\frac{x}{n-1}-2 z+4\right),
\end{equation}

\begin{equation}
	\frac{ds}{du}=\frac{\Delta  z}{n}+s (-x)-2 s z+4 s,
\end{equation}

\begin{equation}
	\label{doi}
	\frac{du}{dN}=u \left(\frac{\Delta  m z}{n s}+2 (m-1) (z-2)-x\right).
\end{equation}

The acceleration equation \eqref{acceleration} can be expressed in the following way:
\onecolumngrid
\begin{equation}
	\Delta = \frac{s \left(6 n z \left((2 m (4 m-5)+3) (n-1) u+(n-2) x^2+2 (n-1) x-n+1\right)-6 (n-1) z^2 (2 m (2 m-1) n u+n-3)+n (n-1)^2 \varsigma\right)}{12 m^2 (n-1) u z^2},
\end{equation}
\twocolumngrid
obtaining a direct relation between $\Delta$ and $\varsigma$. An additional relation between these two non--independent variables is obtained by direct differentiation of the Friedmann constraint equation \eqref{frconstraint}, i.e. 
\begin{multline}
	\frac{n (6 z ((n-2) x^2-n x+2 (n-1) (z-2)+x)+(n-1)^2 \varsigma)}{6 (n-1) z^2}
	\\=\Delta +m u (-\frac{2 n (z-2) (2 m (z-2)-2 z+1)}{z}
	\\+\frac{\Delta  (-2 m (z-2)-1)}{s}-\frac{2 n x}{n-1}).
\end{multline}
In this approach we can close the autonomous system of equations, obtaining specific relations for the non--independent variables $\varsigma$ and $\Delta$. The final form of the autonomous system is the following:

\onecolumngrid 
\begin{equation}
	\frac{dz}{dN}=-\frac{z \left(n u (2 m z-4 m+1)+2 n^2 z-4 n^2-n s-3 n z+5 n+z\right)}{(n-1) n},
\end{equation}

\begin{equation}
	\frac{ds}{du}=s(2 m u z-4 m u+\frac{z}{n}-s+u-3 z+5+ \frac{A}{B}) ,
\end{equation}

\begin{equation}
	A=n s (2 m u z-3 n+3)-2 m u (z^2 (2 m n u-(n-1) (2 (m-1) n+1))+n z (6 m n-4 m u-6 m-7 n+u+8)-(4 m-3) (n-1) n),
\end{equation}

\begin{equation}
	B=(n-1) n (m u (-2 m z+2 m-1)+s),
\end{equation}

\begin{equation}
	\frac{du}{dN}=u(2 m u z-4 m u+2 (m-1) (z-2)+\frac{z}{n}-s+u-z+1+\frac{C}{D}),
\end{equation}

\begin{equation}
	C=m (n s (2 m u z-3 n+3)-2 m u (z^2 (2 m n u-(n-1) (2 (m-1) n+1))+n z (6 m n-4 m u-6 m-7 n+u+8)-(4 m-3) (n-1) n)),
\end{equation}

\begin{equation}
	D=(n-1) n (m u (-2 m z+2 m-1)+s).
\end{equation}

\twocolumngrid
The critical points for the autonomous system are obtained by setting the r.h.s. of the equations \eqref{unu}--\eqref{doi} to zero. In this case we have obtained the following critical points, associated to different cosmological solutions which are attached to various epochs in the evolution of the Universe.
\par 
The first critical point investigated in our analysis is located in the phase space structure at the following coordinates:
\begin{equation}
	\mathcal{Q}_1=\big[z=0, s=2, u=0 \big],
\end{equation}
describing a radiation era $(w_{tot}=\frac{1}{3})$ where the dynamics in influenced mainly by the matter component through the $s$ variable. The corresponding eigenvalues are the following:
\begin{equation}
	\big[-2,3-7 m,\frac{4 n-3}{n-1} \big].
\end{equation}
It can be seen that this epoch can be either stable or saddle, depending on the coupling parameters $n$ and $m$. 
\par
\begin{figure}[tb]
	\centering
	\includegraphics[width=0.4\textwidth]{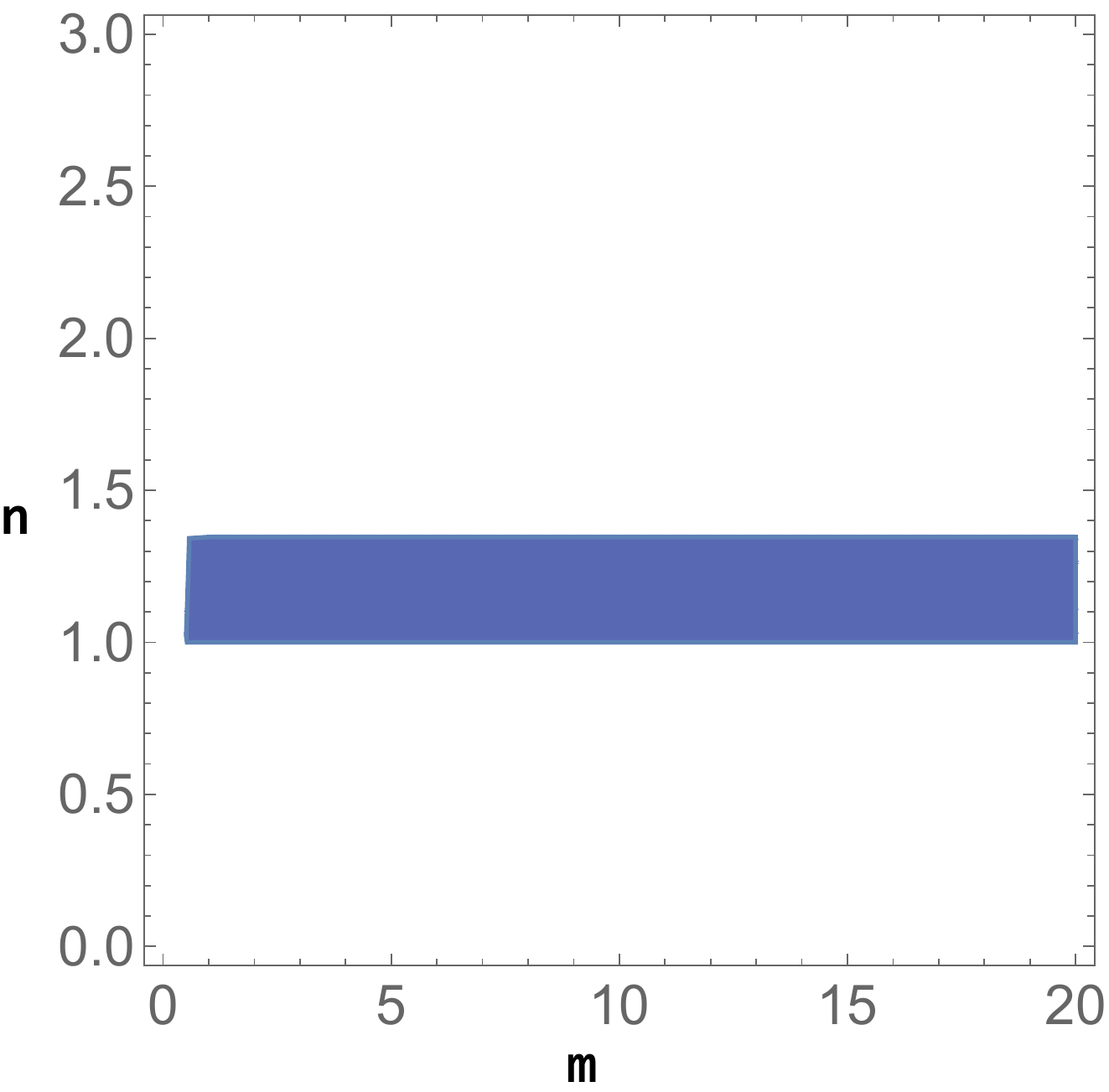}
	\caption{A specific region where the critical point $\mathcal{Q}_2$ is stable and physically viable, with $s \in [0,1]$.}
	\label{fig:punctQ2stabilitate}
\end{figure}

\begin{figure}[tb]
	\centering
	\includegraphics[width=0.4\textwidth]{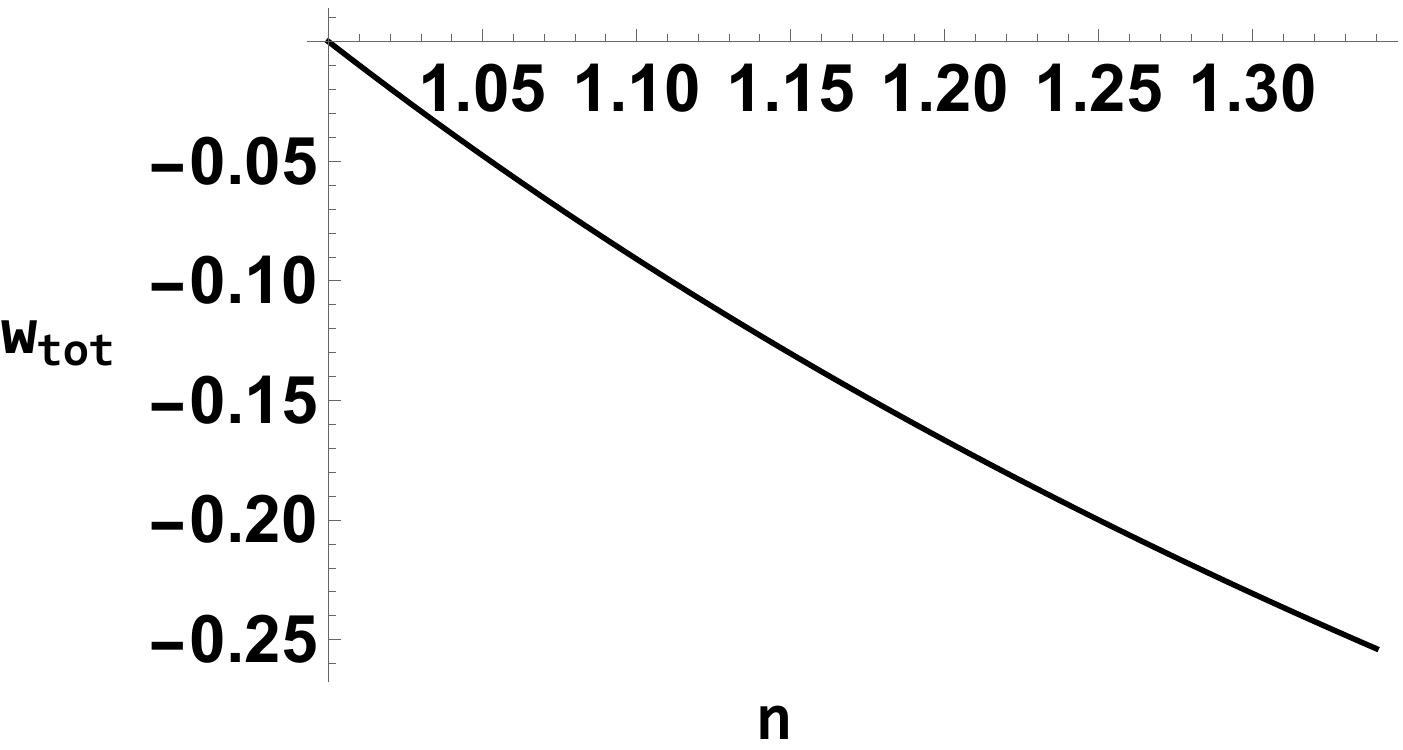}
	\caption{The variation of the total equation of state for the critical point $\mathcal{Q}_2$.}
	\label{fig:punctQ2weff}
\end{figure}

 
The second critical point can be found at the following coordinates: 
\begin{equation}
	\mathcal{Q}_2=\big[z=2-\frac{3}{2 n}, s=\frac{(13-8 n) n-3}{2 n^2}, u=0 \big],
\end{equation}
being influenced by the value of the curvature coupling, without any influence from the matter--geometry interplay. The total equation of state $(w_{tot}=\frac{1}{n}-1)$ can mimic an accelerated expansion era by fine--tuning the value of the curvature coupling parameter $n$. From a dynamical point of view we have obtained the following eigenvalues:
\begin{multline}
	\big[3-\frac{3 m (n+1)}{n},
	\\ \frac{\pm \sqrt{256 n^4-864 n^3+1025 n^2-498 n+81}-3 n+3}{4 (n-1) n} \big].
\end{multline}
From a physical point of view the value of the effective matter density parameter should be real and positive, $s>0$. We have showed in Fig.~\ref{fig:punctQ2stabilitate} a specific region where $\mathcal{Q}_2$ is a stable node where all of the eigenvalues have negative real parts, and the effective matter density parameter is restricted to the $[0,1]$ interval. In Fig.~\ref{fig:punctQ2weff} we see that the value of the total equation of state depends on the scalar curvature coupling parameter. We can note that the main physical features for the critical points $\mathcal{Q}_1$ and $\mathcal{Q}_2$ are not influenced by the matter--geometry invariant, these solutions can be found in different earlier studies, being specific for the $f(R)$ component in the action \cite{Bahamonde:2017ize}. For these solutions the matter--geometry invariant affects the physical features through the dynamics, influencing the values of the corresponding eigenvalues.   
\par 
The third critical point represents a geometrical dark energy solution with the dynamics driven mainly by the matter--geometry coupling invariant, 
\begin{equation}
	\mathcal{Q}_3=\big[z=0, s=0, u=\frac{8 m-5}{8 m^2-6 m+1} \big].
\end{equation}
The solution corresponds to a radiation epoch $(w_{tot}=\frac{1}{3})$, having the following eigenvalues:
\begin{equation}
	\big[\frac{6-10 m}{1-2 m},\frac{5-8 m}{2 m-1},\frac{m (8 n-2)-4 n}{(2 m-1) (n-1)}   \big].
\end{equation}
By proper fine--tuning we can obtain viable restrictions for the constant parameters such that this solution has a saddle dynamic. For example, if we take into account the following restriction,
\begin{equation}
	m<\frac{1}{2}\lor m>\frac{5}{8},
\end{equation}
we obtain a saddle behavior associated to the radiation era.
\par
\begin{figure}[tb]
	\centering
	\includegraphics[width=0.4\textwidth]{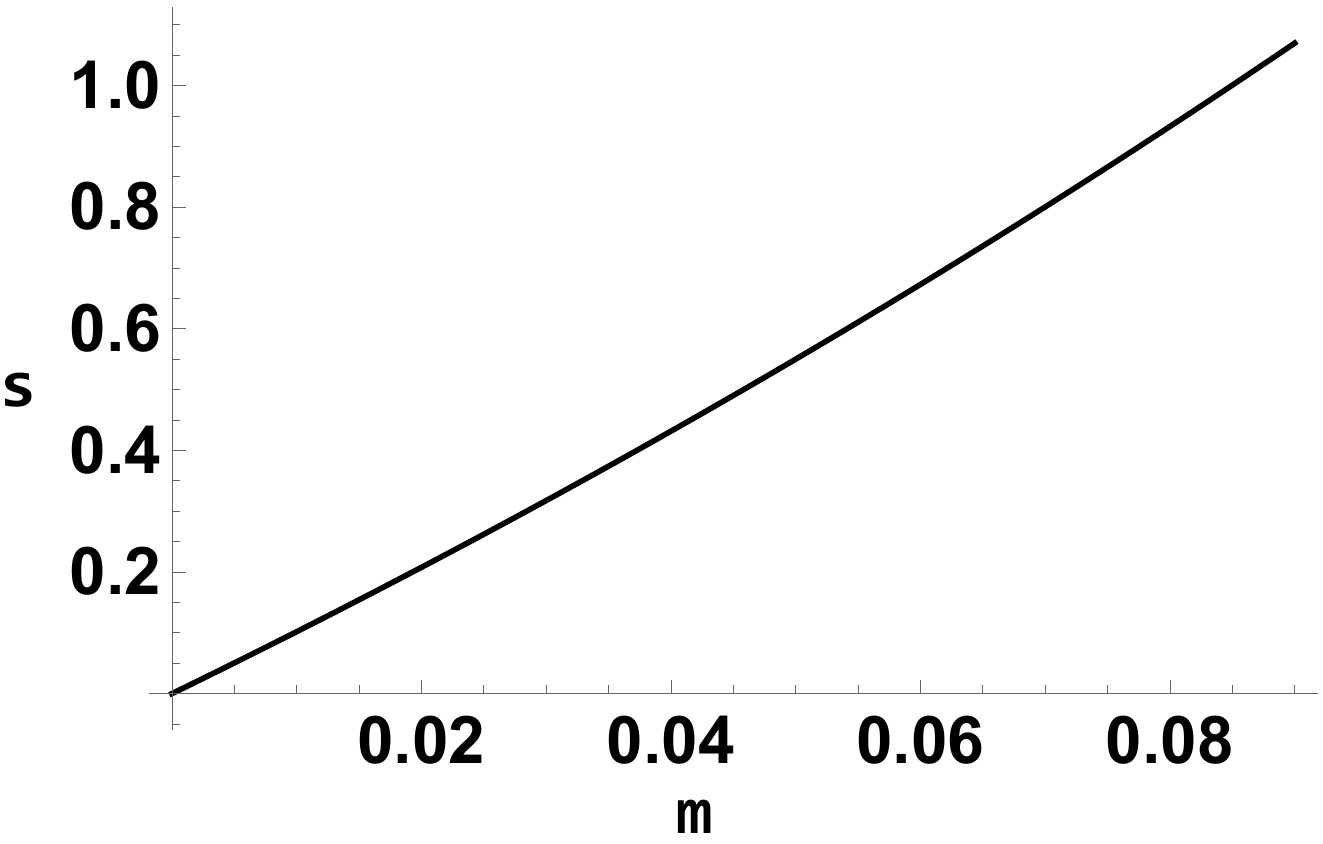}
	\caption{The variation of the $s$ variable for the cosmological solution $\mathcal{Q}_4$.}
	\label{fig:punctQ4s}
\end{figure}

 
Next, the solution $\mathcal{Q}_4$ can be found in the phase space structure at the following coordinates:
\begin{multline}
	\mathcal{Q}_4=\big[z=0, s=-\frac{2 m (5 m-3) (9 m-5)}{(m-1) (m (18 m-13)+3)},
	\\ u=\frac{27 m-9}{m (18 m-13)+3}+\frac{2}{m-1} \big],
\end{multline}
 having a similar behavior as $\mathcal{Q}_3$ $(w_{tot}=\frac{1}{3})$. In this case the effective matter density parameter is affected by the matter--geometry interplay, without any influence by the curvature coupling. We have obtained the following specific eigenvalues:
 \begin{equation}
 	\big[-\frac{2 \left(35 m^2-36 m+9\right)}{14 m^2-23 m+9},\frac{5-9 m}{m-1},\frac{4 (m n+m-n)}{(m-1) (n-1)} \big].
 \end{equation}
As in the previous case, by proper fine--tuning we can identify specific regions where the behavior corresponds to a saddle dynamics, compatible to the known history of the Universe. In Fig.~\ref{fig:punctQ4s} we plot the variation of the matter density parameter for different values of $m$, the auxiliary variable associated to the interplay between matter and geometry. 
\par
\begin{figure}[tb]
	\centering
	\includegraphics[width=0.4\textwidth]{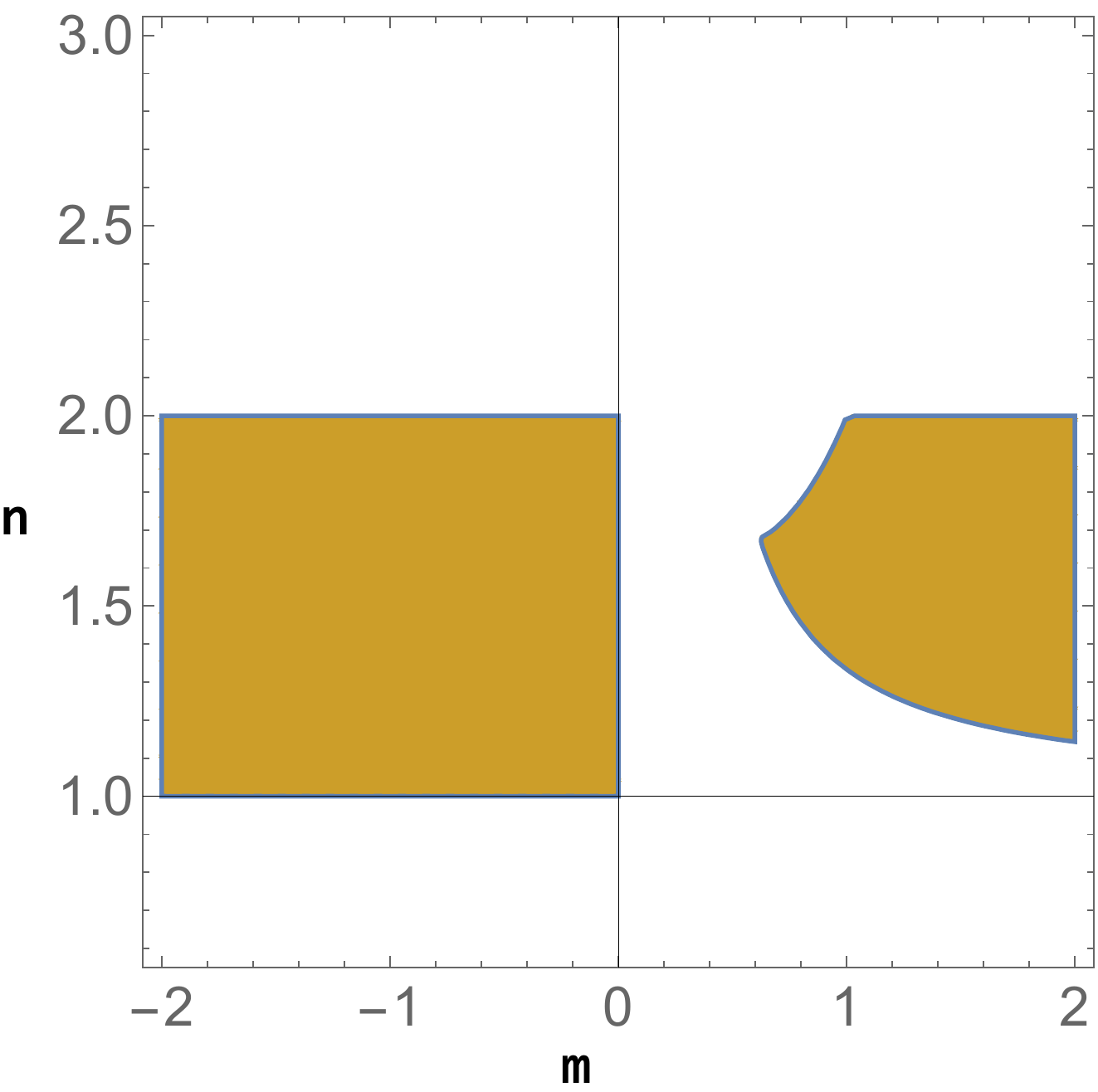}
	\caption{A region where the cosmological solution $\mathcal{Q}_5$ is physically viable and stable $s \in [0,1]$.}
	\label{fig:punctQ5stabilitate}
\end{figure}

\begin{figure}[tb]
	\centering
	\includegraphics[width=0.4\textwidth]{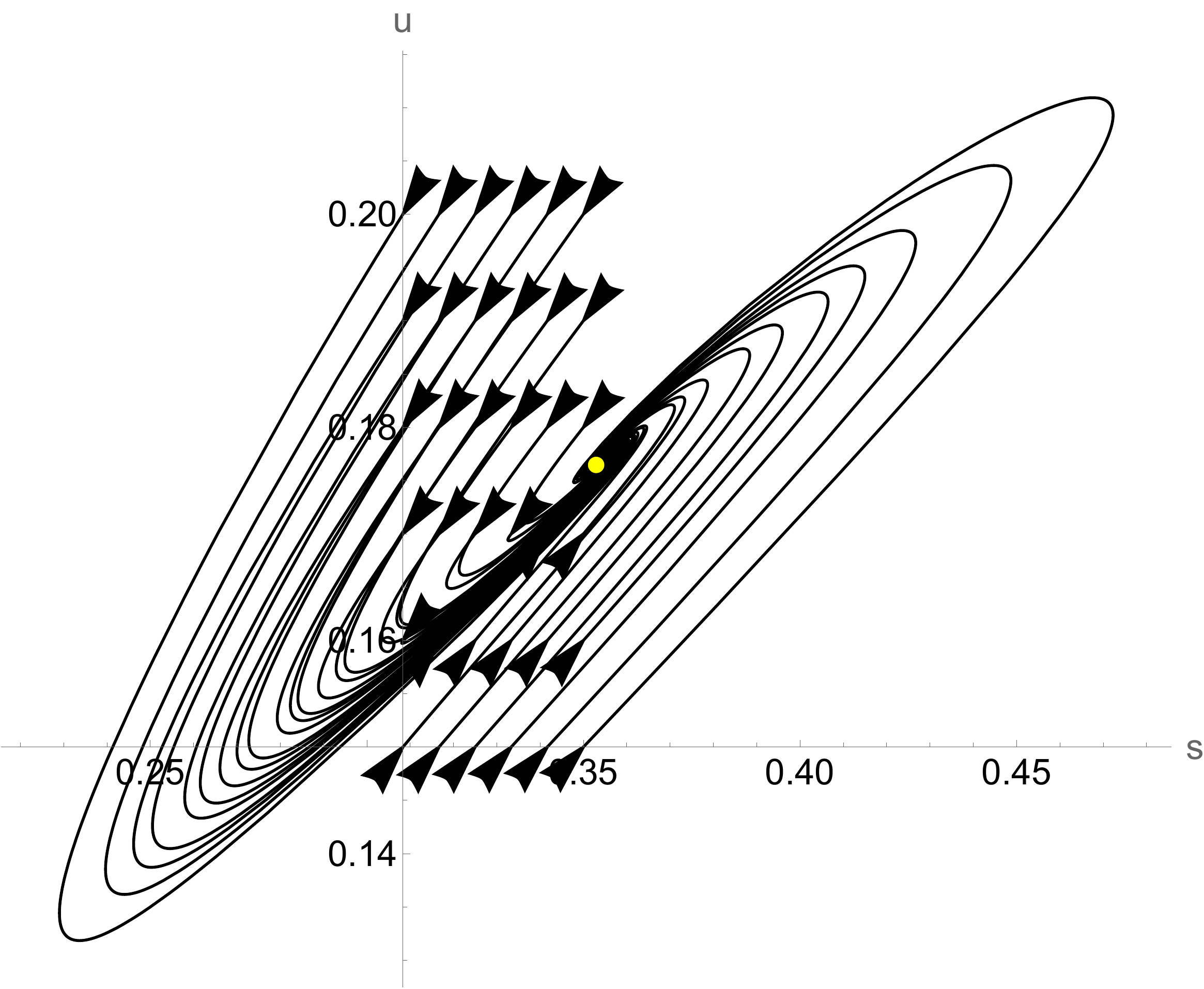}
	\caption{The evolution in the phase space towards the de--Sitter solution $\mathcal{Q}_5$ for different initial conditions in the attractor basin.}
	\label{fig:punctQ5numeric}
\end{figure}

 
For the critical point $\mathcal{Q}_5$ we have the following coordinates:
\begin{equation}
	\mathcal{Q}_5=\big[z=2, s=-\frac{2 m (n-2)}{(2 m-1) n},u=-\frac{2-n}{n-2 m n} \big],
\end{equation}
a particular solution corresponding to a de--Sitter epoch, where the dynamics corresponds to a cosmological constant, $(w_{tot}=-1)$. The formulas associated to the eigenvalues are too complicated to be displayed in the manuscript. As an alternative, we show in Fig.~\ref{fig:punctQ5stabilitate} a particular region of interest where the cosmological solution is stable and physically viable, taking into consideration the following restriction: $s \in [0,1]$. The numerical evolution towards $\mathcal{Q}_5$ solution is presented in Fig.~\ref{fig:punctQ5numeric}, showing the compatibility between the analytical analysis and the numerical approach $(m=1, n=1.7)$.  
\par
\begin{figure}[tb]
	\centering
	\includegraphics[width=0.4\textwidth]{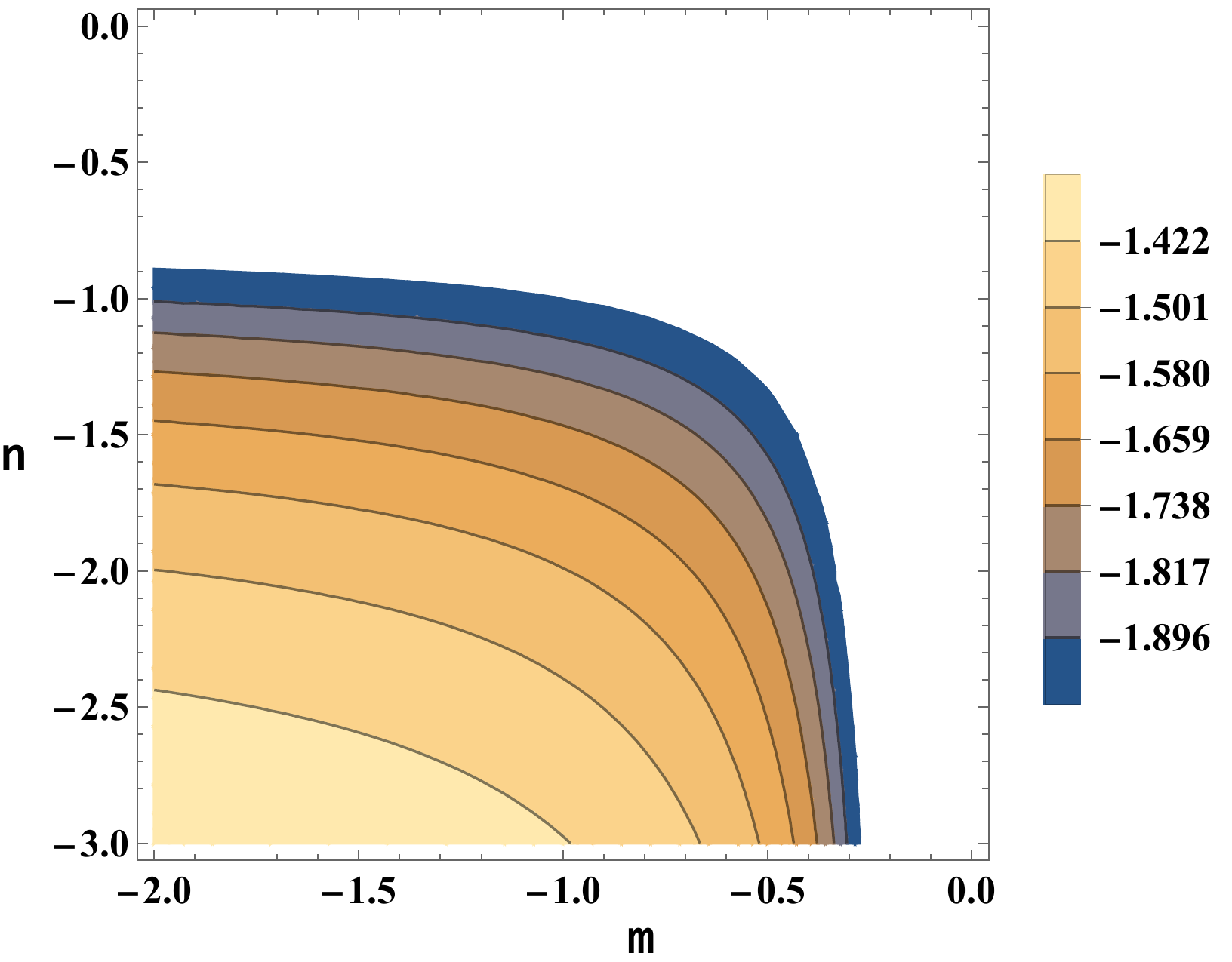}
	\includegraphics[width=0.4\textwidth]{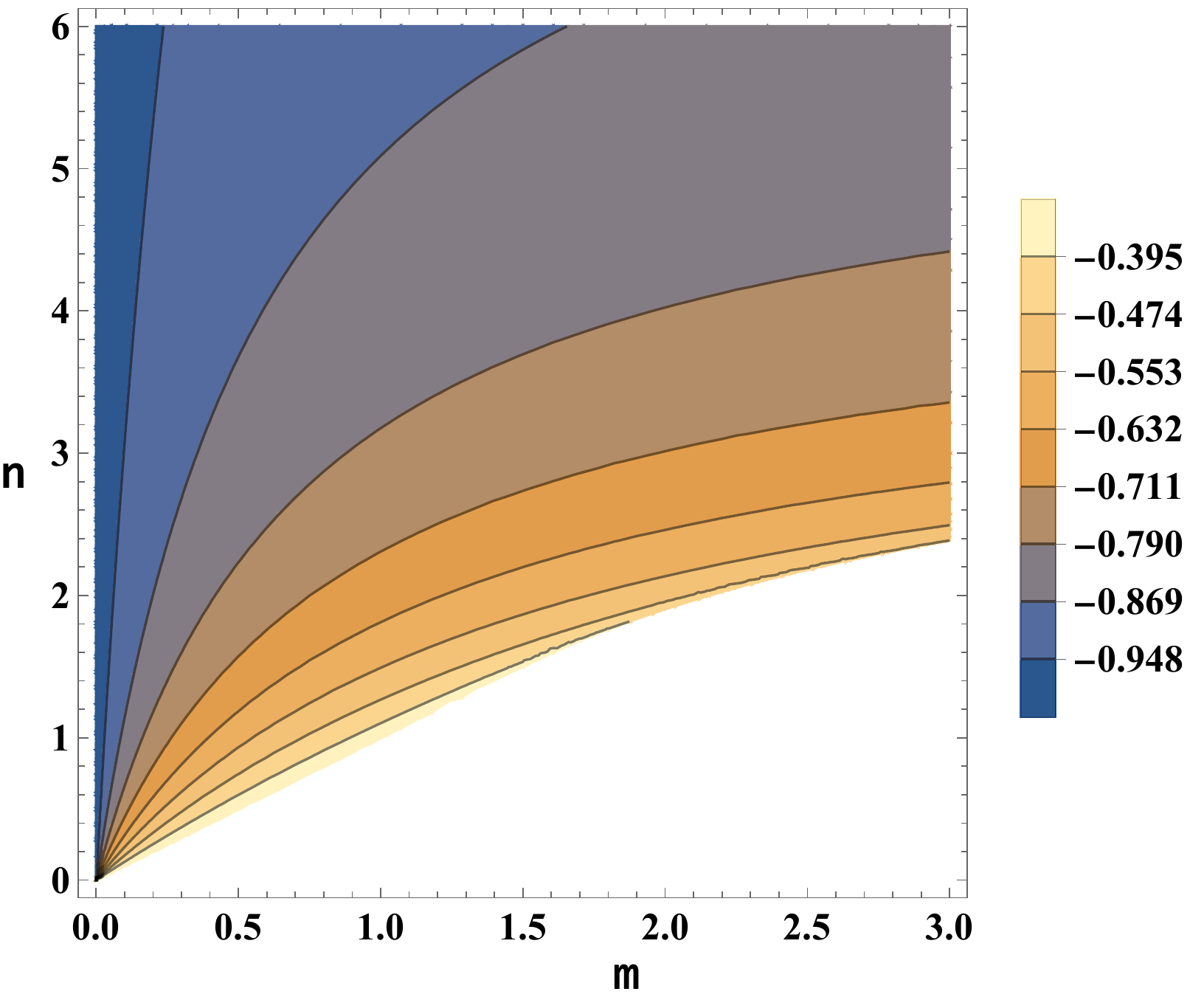}
	\caption{The variation of the total equation of state for the $\mathcal{Q}_6^{+}$ curvature--matter--geometry solution. }
	\label{fig:punctQ6weff}
\end{figure}

\begin{figure}[tb]
	\centering
	\includegraphics[width=0.4\textwidth]{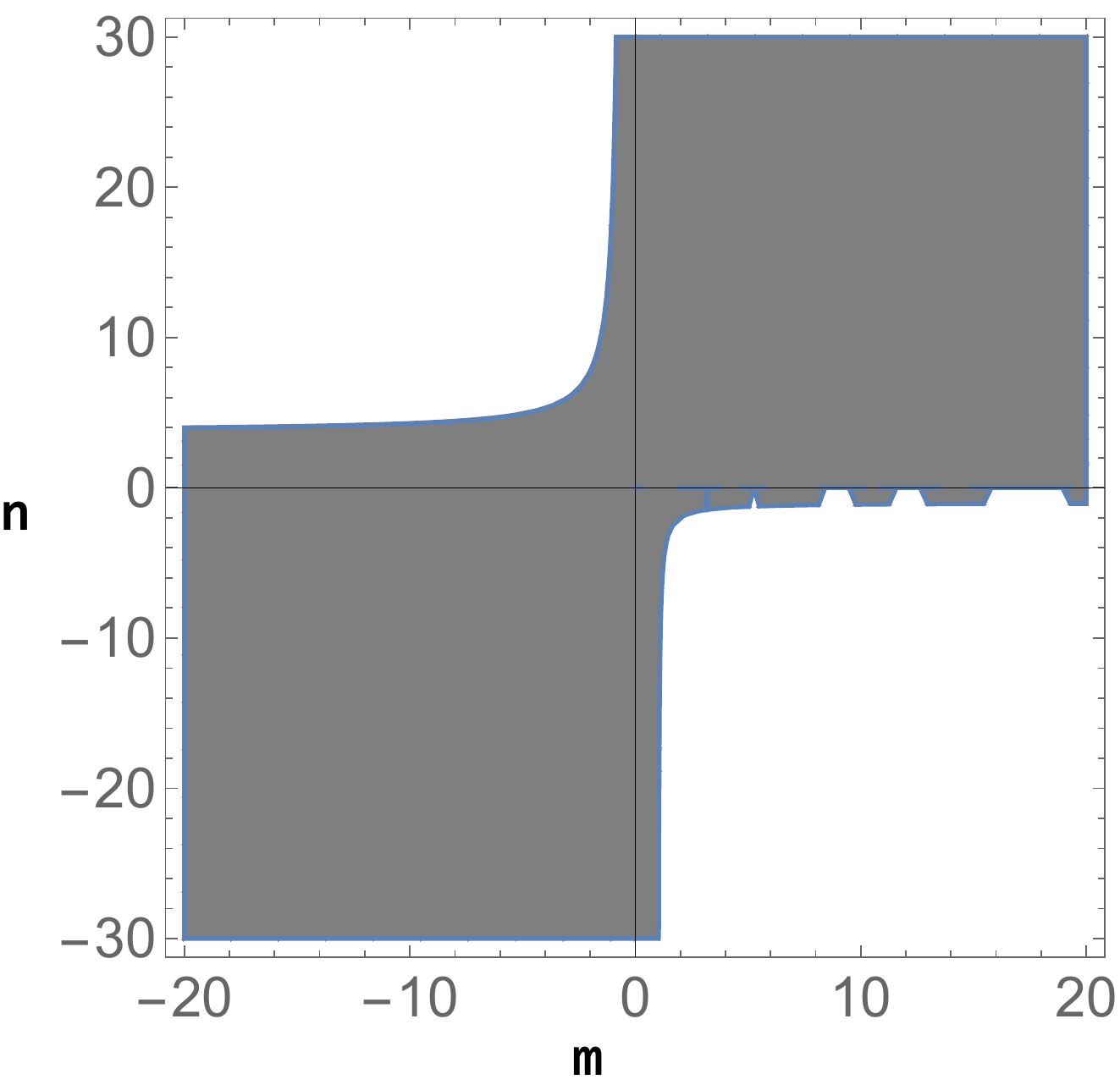}
	\caption{A region where the cosmological solution $\mathcal{Q}_6^{+}$ has a saddle dynamical behavior.}
	\label{fig:punctQ6saddle}
\end{figure}

 
The last class of critical points can be found at the following coordinates:
\onecolumngrid
\begin{multline}
	\mathcal{Q}_6^{\pm}=\big[z=\frac{\pm\sqrt{4 m^2+4 (1-4 m) m n+(2 m n+n)^2}+m (6 n-2)-n}{4 m n}, s=0,
	\\u=\frac{-28 m^2 n^2\mp\left(2 m (n-2)+2 n^2-3 n+1\right) \sqrt{4 m^2 ((n-4) n+1)+4 m n (n+1)+n^2}+\mathcal{R}}{8 m^2 n (3 m-n-1)} \big],
\end{multline}
\twocolumngrid 
with $\mathcal{R}=40 m^2 n-4 m^2+4 m n^3-8 m n+2 m+2 n^3-3 n^2+n$. Since the two solutions are very similar, we shall discuss only the $\mathcal{Q}_6^{+}$ case. As can be noted, the matter--geometry coupling and scalar curvature component are influencing the location in the phase space structure and the corresponding dynamical features. This epoch can be regarded as a curvature--matter--geometry solution, where the geometrical--matter components completely dominates in terms of effective density parameters. For this point, we have obtained the following total equation of state,
\onecolumngrid 
\begin{equation}
	w_{tot}=\frac{-\sqrt{4 m^2+4 (1-4 m) m n+(2 m n+n)^2}-4 m n+2 m+n}{6 m n},
\end{equation} 
\twocolumngrid
with influences from both the scalar curvature coupling and the matter--geometry invariant, respectively. By fine--tuning these parameters, we can obtain different classes of cosmological eras. For example, if we set $m=1$ and $n=\frac{2}{3}$ we obtain a matter dominated epoch, whereas for $m=2$ and $n=\frac{1}{3}$ we get a radiation behavior. In Figs.~\ref{fig:punctQ6weff} we plot the variation of the total equation of state for different values of the $m$ and $n$ parameters. We can observe that by fine--tuning we can  obtain different values, corresponding to de--Sitter, quintessence, and phantom regimes. The final analytical expressions for the corresponding eigenvalues are very cumbersome and are not displayed in the manuscript. Instead, we focus on displaying in Fig.~\ref{fig:punctQ6saddle} a non--exclusive possible interval where the cosmological solution is represented by a saddle critical point.

\section{The phase space analysis for the power--law representation}
\label{sec:apatra} 
Next, we study the phase space structure for the case where the geometrical coupling function extends the fundamental Einsten--Hilbert action, $f(R,\phi)=\frac{R}{2}+g_0 \phi^{\alpha}$, considering a power law representation for the matter--geometry invariant. In this case one needs to introduce the following dimension--less variables:

\begin{equation}
	s=\frac{\rho_m}{3 H^2},
\end{equation}
	
\begin{equation}
	x=\frac{g(\phi)}{3 H^2},
\end{equation}
	
\begin{equation}
	y=\frac{\dot{H}}{H^2}.
\end{equation}

In terms of dimension--less variables, the Friedmann constraint equation \eqref{frconstraint} becomes:
\begin{equation}
	-1+s-x(2 \alpha  y+1)=0, 
\end{equation}
reducing the dimension of the phase space with one unit. Hence, the final independent dimension--less variables are: $\{s,x\}$.
Then, the associated dynamical equations have the following form:
\begin{equation}
	\frac{ds}{dN}=-2 s y + \mathcal{L},
\end{equation}

\begin{equation}
	\frac{dx}{dN}=\frac{\alpha  x \mathcal{L}}{s}+2 \alpha  x y-2 x y,
\end{equation}

where the additional non--independent variable is defined as: $\mathcal{L}=\frac{\dot{\rho_m}}{3 H^3}$. The specific form of the acceleration equation \eqref{acceleration} can be used in order to extract the non--independent variable, 
\onecolumngrid 

\begin{equation}
	\mathcal{L}=-\frac{s ((2 \alpha -1) x (2 \alpha  y+3)-2 y-3)}{2 \alpha ^2 x}.
\end{equation}

\twocolumngrid

Hence, we obtain the final form of the autonomous system:

\onecolumngrid

\begin{equation}
	\frac{ds}{dN}=\frac{s \left(x \left(4 \alpha ^2+2 \alpha +(1-4 \alpha ) \alpha  s-1\right)+s-2 (\alpha -1) \alpha  x^2-1\right)}{2 \alpha ^3 x^2},
\end{equation}

\begin{equation}
	\frac{dx}{dN}=\frac{-x (-4 \alpha +\alpha  s+1)+s+2 \alpha  (2-3 \alpha ) x^2-1}{2 \alpha ^2 x}.
\end{equation} 

\twocolumngrid
As in the previous case, we have identified the following critical points by analyzing the r.h.s. of the autonomous system. 
\par
\begin{figure}
	\centering
	\includegraphics[width=0.4\textwidth]{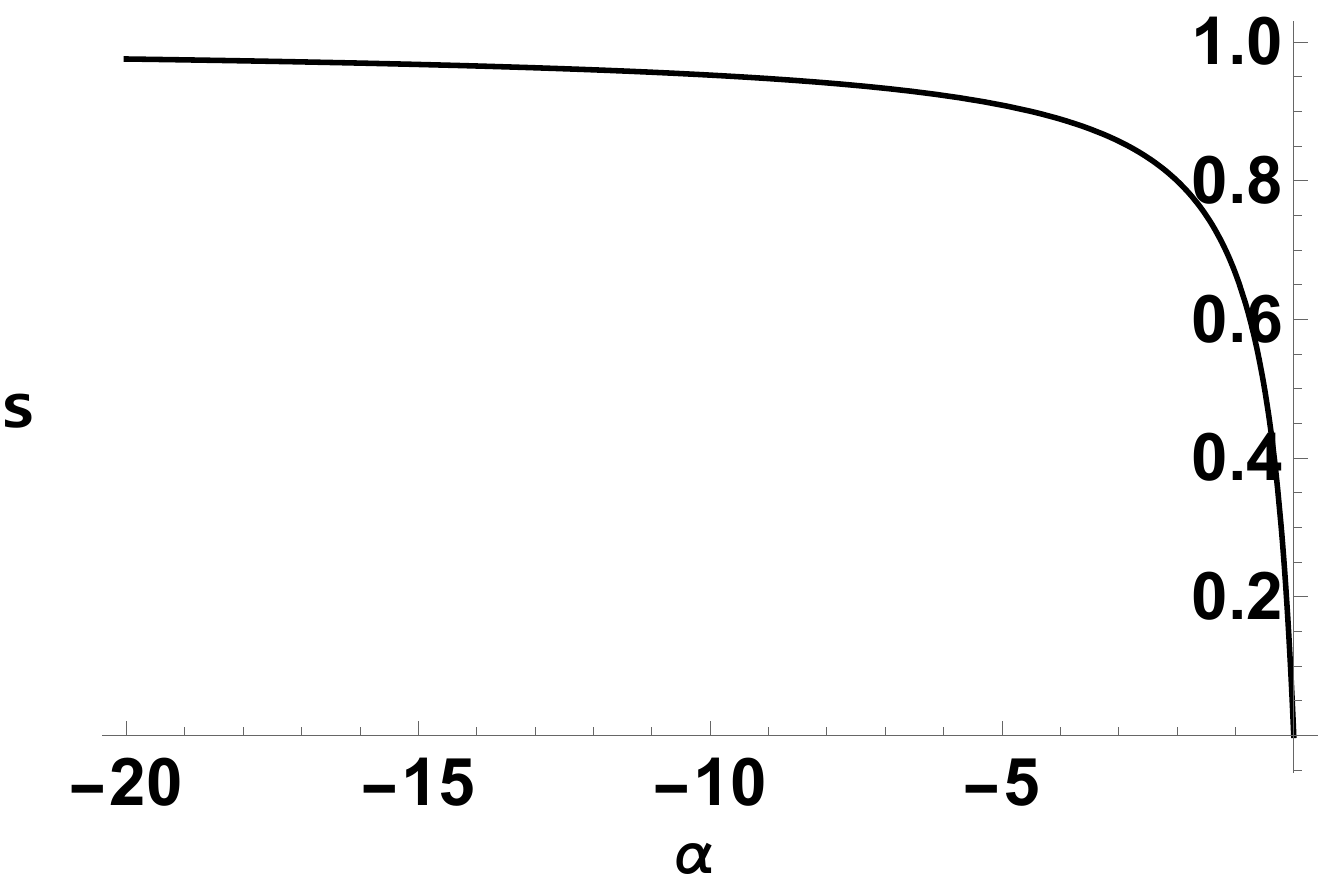}
	\caption{The value of the matter density parameter for the de--Sitter critical point $\mathcal{U}_1$.}
	\label{fig:punctU1s}
\end{figure}

 
The first critical points is located at the following coordinates:
\begin{equation}
	\mathcal{U}_1=\Bigg[s=\frac{2 \alpha }{2 \alpha -1}, x=\frac{1}{2 \alpha -1} \Bigg],
\end{equation}
 representing a de--Sitter epoch $(w_{tot}=-1)$ having the following eigenvalues:
 \begin{equation}
 	\Bigg[\frac{2 \alpha  (4-7 \alpha )\pm \sqrt{4 (\alpha -1) \alpha  ((\alpha -7) \alpha +4)+1}-1}{4 \alpha ^2} \Bigg].
 \end{equation}
For this point we have displayed in Fig.~\ref{fig:punctU1s} the value of the $s$ variable in the case where $\alpha$ is negative. The $s$ variable represents the matter density parameter and have to be positive from a physical point of view. As can be noted, this critical point is associated to an era where the geometrical coupling function acts as a cosmological constant. From a dynamical perspective in the case where $\alpha$ is negative the point represents an attractor, with negative eigenvalues.

\par
\begin{figure}
	\centering
	\includegraphics[width=0.4\textwidth]{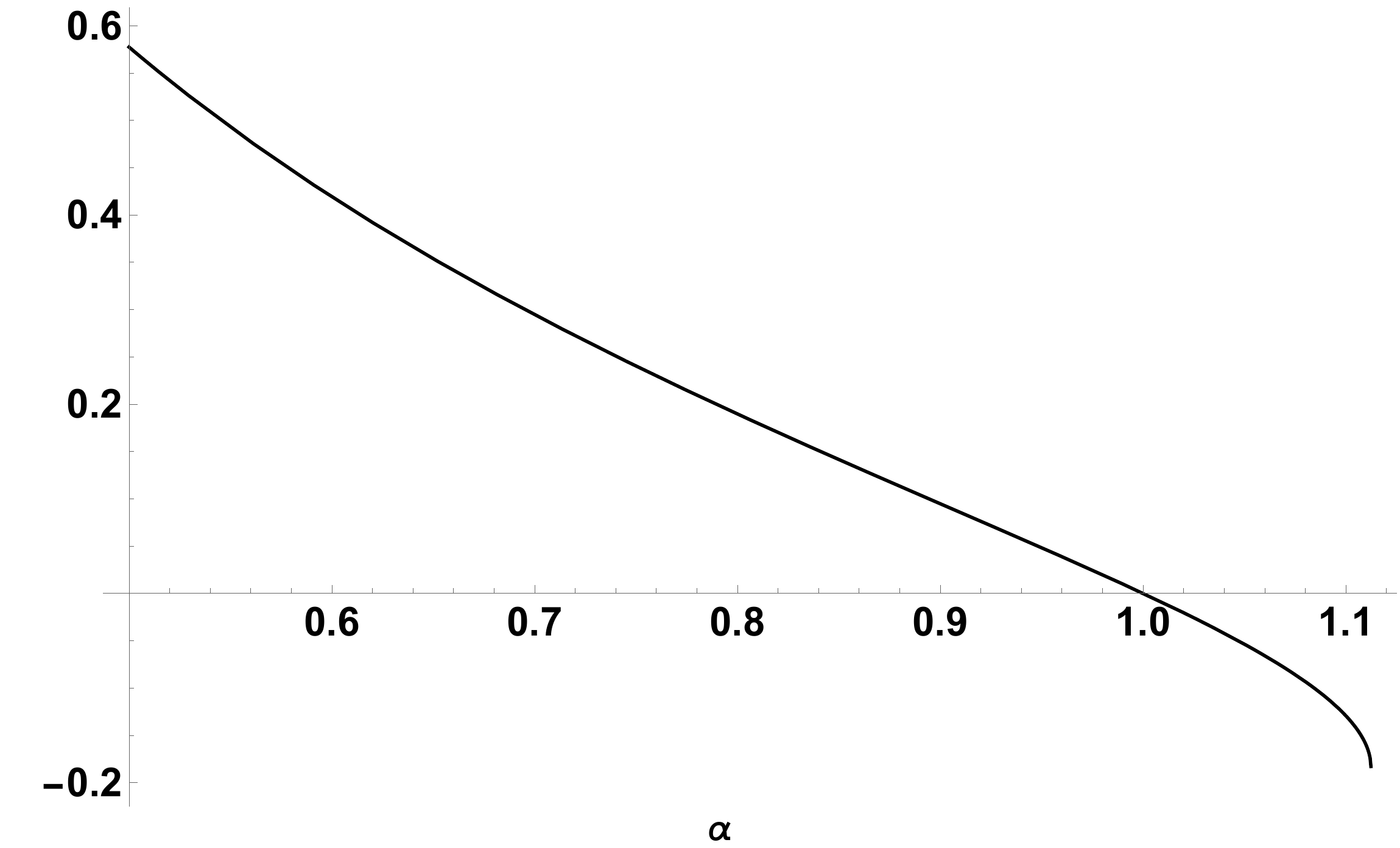}
	\includegraphics[width=0.4\textwidth]{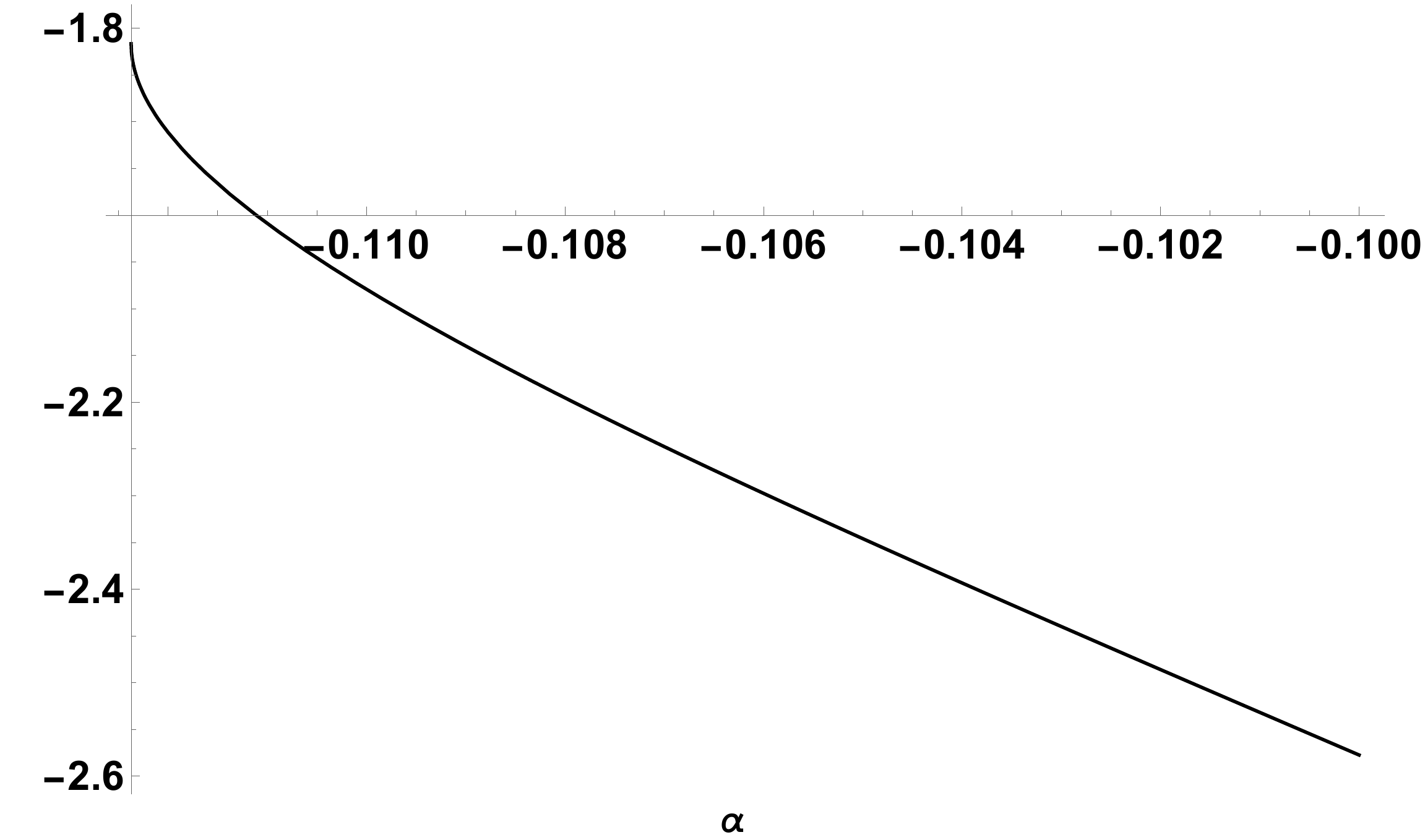}
	\caption{The total equation of state for the $\mathcal{U}_2^{+}$ critical point.}
	\label{fig:punctU2weff}
\end{figure}

 
The second critical point represents an epoch located in the phase space at:
\begin{equation}
	\mathcal{U}_2^{\pm}=\Bigg[s=0, x=\frac{\pm \sqrt{-8 \alpha ^2+8 \alpha +1}-4 \alpha +1}{8 \alpha -12 \alpha ^2} \Bigg],
\end{equation}
an epoch which can describe the accelerated expansion with the effective equation of state:
\begin{equation}
	w_{tot}=\frac{-2 \alpha \pm \sqrt{1-8 (\alpha -1) \alpha }+1}{6 \alpha },
\end{equation}
being sensitive to the value of the geometrical coupling parameter $\alpha$. We note that the values of the matter density parameter is zero, a feature where the geometrical--matter coupling plays a fundamental role in the effective dynamics at the background level. In what follows we briefly describe the 	$\mathcal{U}_2^{+}$ solution. The dynamics in a specific region for various values of the coupling parameter $\alpha$ is represented in Fig.~\ref{fig:punctU2weff}. As can be seen, this critical point can explain various staged in the history of our Universe, the super--acceleration, the radiation $(\alpha =\frac{2}{3})$ and the matter domination $(\alpha =1)$ by fine--tuning. The accelerated expansion ($w_{tot}<-\frac{1}{3}$) is obtained in the following interval: $(\frac{1}{4} \left(2-\sqrt{6}\right)\leq \alpha <0)$. From a dynamical perspective for the $\mathcal{U}_2^{+}$ solution we have obtained the following eigenvalues:
\onecolumngrid
\begin{multline}
	\Bigg[ \frac{2 (3 \alpha -2) \left(8 \alpha ^2+4 \left(\sqrt{-8 \alpha ^2+8 \alpha +1}-2\right) \alpha -\sqrt{-8 \alpha ^2+8 \alpha +1}-1\right)}{\alpha  \left(\sqrt{-8 \alpha ^2+8 \alpha +1}-4 \alpha +1\right)^2},
	\\
	\frac{2 (2 \alpha -1) \left(24 \alpha ^3-6 \left(\sqrt{-8 \alpha ^2+8 \alpha +1}+3\right) \alpha ^2+\sqrt{-8 \alpha ^2+8 \alpha +1}+4 \alpha +1\right)}{\alpha ^2 \left(\sqrt{-8 \alpha ^2+8 \alpha +1}-4 \alpha +1\right)^2} \Bigg].
\end{multline}
\twocolumngrid 
Considering the analytical analysis of the previous eigenvalues, the  $\mathcal{U}_2^{+}$ solution is a stable node (with real and negative eigenvalues) in the following interval: $\frac{1}{4} \left(2-\sqrt{6}\right)<\alpha <0$. Finally, we observe that the latter solution can explain a variety of cosmological eras, with a high sensitivity to the values of the geometrical--matter coupling embedded into the $\alpha$ coefficient. For the $\mathcal{U}_2^{-}$ solution we have a similar behavior, noting that $\mathcal{U}_2^{-}$ cannot explain matter and radiation epochs, while the accelerated expansion phenomenon is favored.

\section{The phase space analysis for the exponential model}
\label{sec:acincea} 
\par 
In this section we shall consider the following subclass of functions defined in the following way:
\begin{equation}
	f(R,\phi)=\frac{R}{2}+g_0 e^{\alpha \phi},
\end{equation}
with $f_0$ and $\alpha$ constant parameters. As can be noted, the action is based on the fundamental Einstein--Hilbert action, extended with an exponential representation for the matter--geometry component. For this specific model we have to introduce the following auxiliary variables:
		
\begin{equation}
	s=\frac{\rho_m}{3 H^2},
\end{equation}
		
\begin{equation}
	x=\frac{g(\phi)}{3 H^2},
\end{equation}
		
\begin{equation}
	y=H^2 \dot{H},
\end{equation}
		
\begin{equation}
	z=\frac{1}{H^2}.
\end{equation}
		
Moreover, we add the following non--independent variable 
\begin{equation}
	\mathcal{L}=H \dot{\rho_m},
\end{equation} 
which shall be used in order to form the autonomous system. In these variables, the Friedmann constraint equation \eqref{frconstraint} reduces to the following relation:
	
\begin{equation}
	1+x+s(-1+18 \alpha x y)=0.
\end{equation} 
Then, the autonomous system becomes:
\begin{equation}
	\frac{ds}{dN}=\frac{z^2 \mathcal{L}}{3}-2 s y z^2,
\end{equation}
		
\begin{equation}
	\frac{dx}{dN}=18 \alpha  s x y-2 x y z^2+3 \alpha  x \mathcal{L}, 
\end{equation}

\begin{equation}
	\frac{dz}{dN}=-2 y z^3, 
\end{equation}

while the acceleration equation \eqref{acceleration} is equal to:
		
\onecolumngrid
\begin{equation}
	\frac{3 x \left(108 \alpha ^2 s^2 y+6 \alpha  s \left(y z^2+3 \alpha  \mathcal{L}+3\right)+z^2 (2 \alpha  \mathcal{L}-1)\right)}{z^3}=2 y z+\frac{3}{z}
\end{equation}
\twocolumngrid
Due to the specific form of the acceleration equation \eqref{acceleration}, we can extract the $\mathcal{L}$ variable, remaining with a three dimensional dynamical system $\{s,x,z\}$. As in the previous cases, we determine the associated critical points by analyzing the right hand side of the autonomous system of differential equations. For the last physical model where the coupling function is represented by an exponential we have identified only one class of cosmological solutions located in the phase space structure at the following coordinates:
		
\begin{equation}
	\Xi_1=\Big[s=\frac{18 \alpha +z^2}{18 \alpha }, x=\frac{z^2}{18 \alpha } \Big].
\end{equation}
		
\begin{figure}[t]
	\centering
	\includegraphics[width=0.4\textwidth]{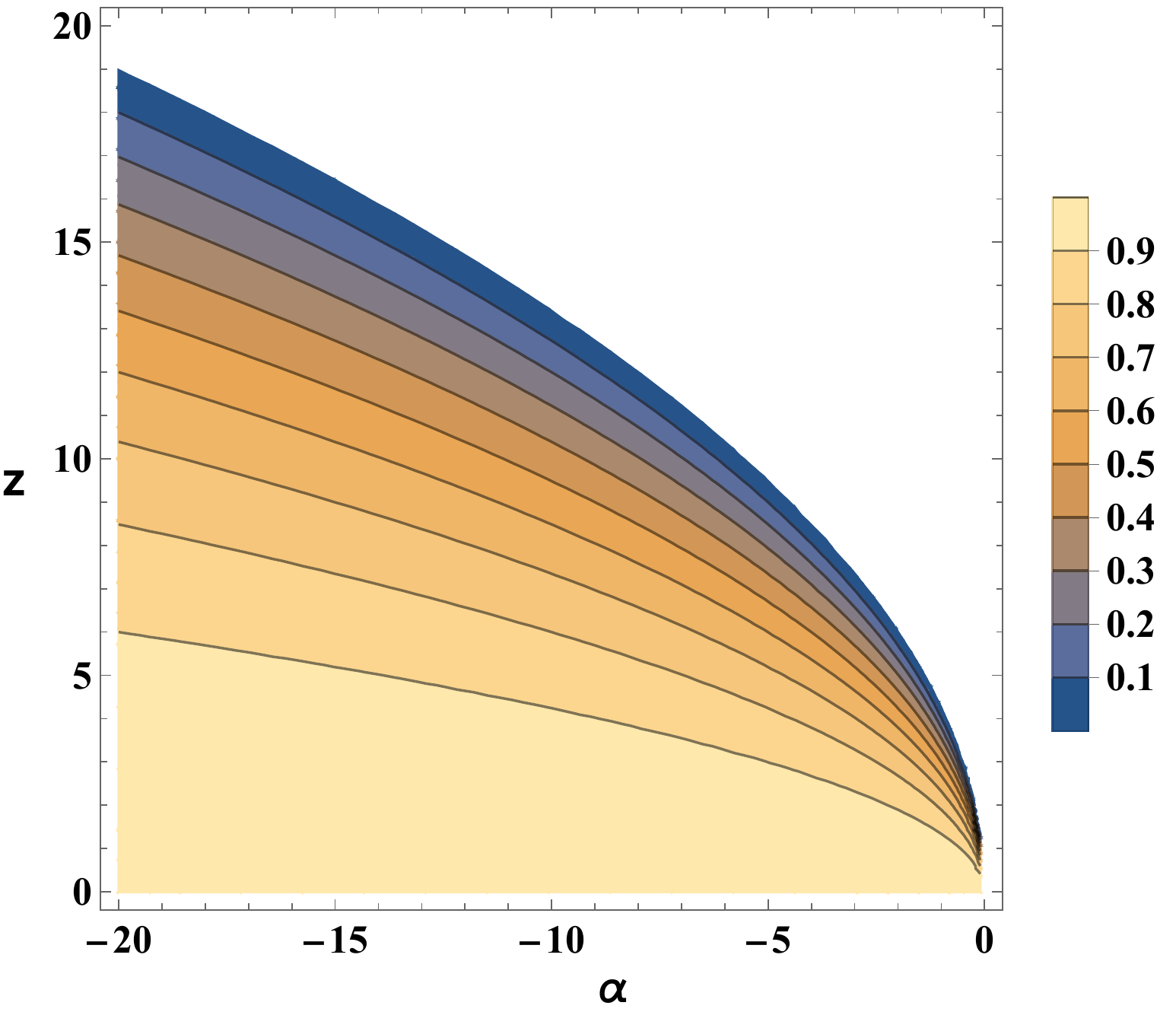}
	\caption{The figure describes a specific region where the de--Sitter critical point $\Xi_1$ is viable, with the matter density parameter in the $[0,1]$ interval.}
	\label{fig:punctxi1s}
\end{figure}

The critical line corresponds to a de--Sitter solution $(w_{tot}=-1)$, describing an epoch where the dynamics induced by the interplay between geometry and matter mimics a cosmological constant behavior. In Fig.~\ref{fig:punctxi1s} we plot a specific region in the $\{z, \alpha \}$ space where the critical line $\Xi_1$ is viable taking into account the existence conditions.

At this cosmological solution we have obtained the following eigenvalues, 
		
\onecolumngrid
\begin{equation} 
	\Bigg[0,-\frac{2268 \alpha ^3+11 \alpha  z^4+360 \alpha ^2 z^2 \pm \sqrt{\alpha ^2 \left(104976 \alpha ^4-23 z^8-720 \alpha  z^6-7128 \alpha ^2 z^4-46656 \alpha ^3 z^2\right)}}{648 \alpha ^3+6 \alpha  z^4+144 \alpha ^2 z^2} \Bigg],
\end{equation}
\twocolumngrid
describing a non--hyperbolic solution which can be saddle in some specific intervals. For example, if we take into account the following restrictions, 
\onecolumngrid 
\begin{equation}
	z\in \mathbb{R}\land z\neq 0\land -\frac{z^2}{6}<\alpha <-\frac{z^2}{18},
\end{equation}
\twocolumngrid 
we obtain a de--Siter epoch with a saddle behavior where the dynamics corresponds to a cosmological constant added in the Einstein field equations.

\section{Conclusions}
\label{sec:concluzii}
\par 
In the present paper we have investigated new cosmological models based on specific couplings between the matter energy--momentum tensor and the geometrical part, encoded into the Einstein tensor. After deducing the field equations in the case where the matter component is represented by a barotropic fluid without pressure, we have investigated the physical aspects for the corresponding dynamics by adopting the linear stability theory. The linear stability theory represents a particular mathematical approach which is associated to the global dynamics, determining the possible evolution of our Universe in the phase space structure. The first theoretical model investigated is associated to a generic function of the following type: $f(R,\phi)=f_0 R^n+g_0 \phi^m$, with $f_0, g_0, n, m$ constant parameters. In this case the phase space is described by a three dimensional system, having various critical points which are associated to different epochs compatible with the late--time evolution of our Universe. For this model we have obtained the following classes of cosmological eras: de Sitter, radiation, and specific solutions which depend on the coupling parameters $n$ and $m$. These solutions can describe the acceleration epoch and the late--time dynamics in the Universe. For each class of cosmological solutions we have obtained possible constraints of the model parameters from a dynamical perspective, encoded into the specific values of the corresponding eigenvalues. 
\par 
The second cosmological model is associated to the following coupling function, $f(R,\phi)=\frac{R}{2}+g_0 \phi^{\alpha}$, extending the fundamental Einstein--Hilbert action with a power--law model which encodes non--minimal effects from the matter--geometry interplay. In the second case the phase space complexity is reduced, being associated with a second order dynamical system. The cosmological solutions found can describe the de--Sitter epoch and the late--time acceleration. Moreover, for various values of the coupling parameter we can obtain different cosmological solutions, like the radiation or the quintessence/phantom--like dynamics. As in the previous case, for each class of cosmological solutions we have obtained possible dynamical constraints.
\par 
The last cosmological model is described by the following coupling function, $f(R,\phi)=\frac{R}{2}+g_0 e^{\alpha \phi}$, with $g_0, \alpha$ constant parameters. As can be noted, this particular cosmological system extends the fundamental Einstein--Hilbert term with a generic exponential function, encoding specific contractions between the matter energy--momentum tensor and the Einstein tensor. The corresponding phase space is three dimensional, having only one critical line associated to the de--Sitter epoch, explaining the late--time accelerated expansion in the Universe whose dynamics closely follows the dynamics induced by a cosmological constant added to the Einstein--Hilbert term. From a dynamical perspective this particular model can explain only the accelerated expansion and the late--time evolution. Hence, in this specific model the evolution associated to the early time can be explained only by fine--tuning.
\par 
As can be noted, this cosmological model based on the interplay between matter and geometry can explain various dynamical effects in the evolution of our Universe, representing an interesting and viable approach in the modified gravity theories. The present model can be further investigated by adopting different theoretical approaches. For example, one might consider an observational study which in principle can add specific constraints to the particular generic functions analyzed in the present manuscript. In these approaches particular reconstruction methods can in principle be applied, obtaining different physical effects in the late--time evolution. All of these aspects can show the viability and limitations for the present cosmological approaches.   

\section{Acknowledgements}
For the present project we have considered various analytical computations in \textbf{Wolfram Mathematica} \cite{Mathematica} -- \textbf{xAct} \cite{xact}.

\bibliography{article}
\bibliographystyle{apsrev}

\end{document}